%% file: dlsm.tex
\documentclass[10pt,conference,compsocconf,letterpaper]{IEEEtran}
\usepackage{subfigure}
\usepackage{url}
\usepackage{comment}
\usepackage{float}
\usepackage{wrapfig}

\usepackage{tikz}
\newcommand*\circled[1]{\tikz[baseline=(char.base)]{
            \node[shape=circle,draw,inner sep=0.2pt] (char) {#1};}}
            
\begin{document}
\sloppy







\title{Re-enabling high-speed caching for LSM-trees}



\author{Lei Guo$^1$, Dejun Teng$^2$, Rubao Lee$^2$, Feng Chen$^3$, Siyuan Ma$^2$, Xiaodong Zhang$^2$\\
    \\
           $^1$Google Inc. \qquad $^2$The Ohio State University \qquad $^3$Louisiana State University \\
           $^1$leguo@google.com \quad $^2$teng.102@osu.edu, \{liru,masi,zhang\}@cse.ohio-state.edu \quad $^3$fchen@csc.lsu.edu \\
       }

\maketitle

\begin{abstract}
\input{abstract}
\end{abstract}

\input{intro}

\input{background}

\input{motivation}

\input{compaction}

\input{queryprocess}

\input{discuss}

\input{experiments}

\input{related}

\input{conclusion}

\bibliographystyle{abbrv}
\bibliography{hlsm-tree}

\appendices

\input{appendix-data-update-rate}

\end{document}

%% file: abstract.tex
LSM-tree has been widely used in cloud computing systems by Google, Facebook, and Amazon, to achieve high performance for write-intensive workloads.  However, in LSM-tree, random key-value queries can experience long latency and low throughput due to the interference from the compaction, a basic operation in the algorithm, to caching.  LSM-tree relies on frequent compaction operations to merge data into a sorted structure.  After a compaction, the original data are reorganized and written to other locations on the disk.  As a result, the cached data are invalidated since their referencing addresses are changed, causing serious performance degradations.  

We propose dLSM in order to re-enable high-speed caching during intensive writes.  dLSM is an LSM-tree with a compaction buffer on the disk, working as a cushion to minimize the cache invalidation caused by compactions.  The compaction buffer maintains a series of snapshots of the frequently compacted data, which represent a consistent view of the corresponding data in the underlying LSM-tree.  Being updated in a much lower rate than that of compactions, data in the compaction buffer are almost stationary.  In dLSM, an object is referenced by the disk address of the corresponding block either in the compaction buffer for frequently compacted data, or in the underlying LSM-tree for infrequently compacted data.  Thus, hot objects can be effectively kept in the cache without harmful invalidations.  With the help of a small on-disk compaction buffer, dLSM achieves a high query performance by enabling effective caching, while retaining all merits of LSM-tree for write-intensive data processing.  We have implemented dLSM based on LevelDB.  Our evaluations show that with a standard DRAM cache, dLSM can achieve 5--8x performance improvement over LSM with the same cache on HDD storage.  

%% file: intro.tex
\section{Introduction}

With the rise of cloud computing in enterprises and
user-centric Internet services, 
data that are generated and accessed in computing systems 
continue to increase at a high pace. 
There is an increasing need to access user data 
that are created and updated rapidly 
in real time, 
in services like advertisement matching, personalized recommendations, 
social media trend detection, and others. 
A high performance and cost-effective storage infrastructure is highly demanded 
for quality data services in both reads and writes.

\subsection{Problem statement}

\label{sec:intro-problem}

LSM-tree \cite{OCGO96} was originally designed for high throughput transaction systems. 
It writes data to disk sequentially 
and keeps them sorted in multiple levels with merge operations (also known as compactions). 
Avoiding in-place updates, LSM-tree can achieve a high write throughput and 
conduct fast range queries on hard disk drives (HDDs). 
For these merits, LSM-tree has been widely used
in big data systems by industries, such as Bigtable~\cite{Bigtable}, HBase~\cite{HBase}, 
Cassandra~\cite{Cassandra}, and Riak~\cite{BashoRiak},
and is the de facto model for write-intensive data processing. 
However, serving random access based queries over write-intensive data 
with LSM-tree faces a new challenge from buffer caching. 
Caching is a widely used technique 
to accelerate the random read performance of storage systems. 
The principle of caching is to buffer frequently accessed data 
in DRAM memory for quick accesses.
For small objects, data are usually referenced and accessed in blocks stored on disk by the caching system. 
This referencing mechanism works well for update-in-place systems such as B-tree based databases. 
However, the merge operations of LSM-tree frequently reorganize data stored on disk and 
change the location of objects. 
As a result, cached data are invalidated accordingly, and thus
the advantage of buffer caching is significantly reduced. 


\setlength{\intextsep}{2pt}
\setlength{\columnsep}{0.5pc}
\begin{wrapfigure}{L}{0.58\linewidth}
\centering
\includegraphics[width=1\linewidth]{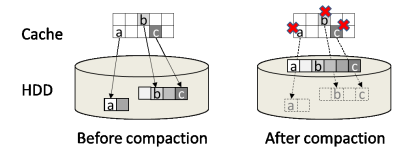}
\vspace{-5mm}
\caption{Cache invalidation due to compaction}
\label{fig:cache-invalidation}
\end{wrapfigure}
The compaction of LSM-tree can be considered as external sorting.  
As shown in Figure \ref{fig:cache-invalidation}, $a$, $b$, and $c$ are frequently requested objects, 
which belong to two sorted tables stored on the disk. 
Originally, the disk blocks containing these objects are kept in the cache. 
However, when the two tables are compacted into a single table, 
the compacted data are written to a new location on the disk. 
Thus, even though the contents of these objects remain unchanged, 
the cached data containing these objects have to be invalidated
since the underlying disk data have been removed. 
When an object is requested again, the system has to load the new data block containing this object from disk. 
With the changes of their referencing addresses, the access information of these objects is also lost. 
Even worse, since compaction writes are conducted in batch, 
the corresponding cache invalidations and data reloading often occur in bursts, 
causing significant performance churns \cite{AK15,cassandra-compact}. 

\setlength{\columnsep}{2pc}
\setlength{\intextsep}{10pt plus 4pt minus 3pt}



\begin{figure}[t]
\centerline{
\subfigure[Read only workload]{\includegraphics[width=0.5\linewidth]{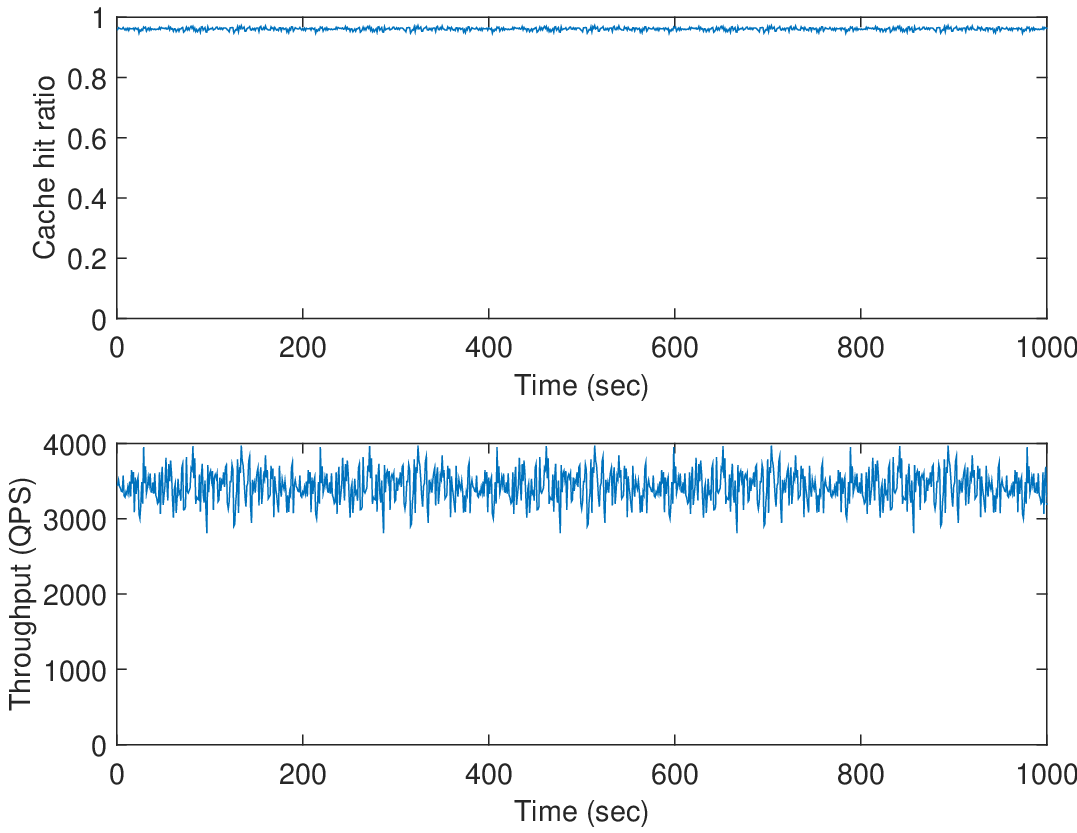}\label{fig:demo-readonly}}
\hspace{0.2mm}
\subfigure[Read and write workload]{\includegraphics[width=0.5\linewidth]{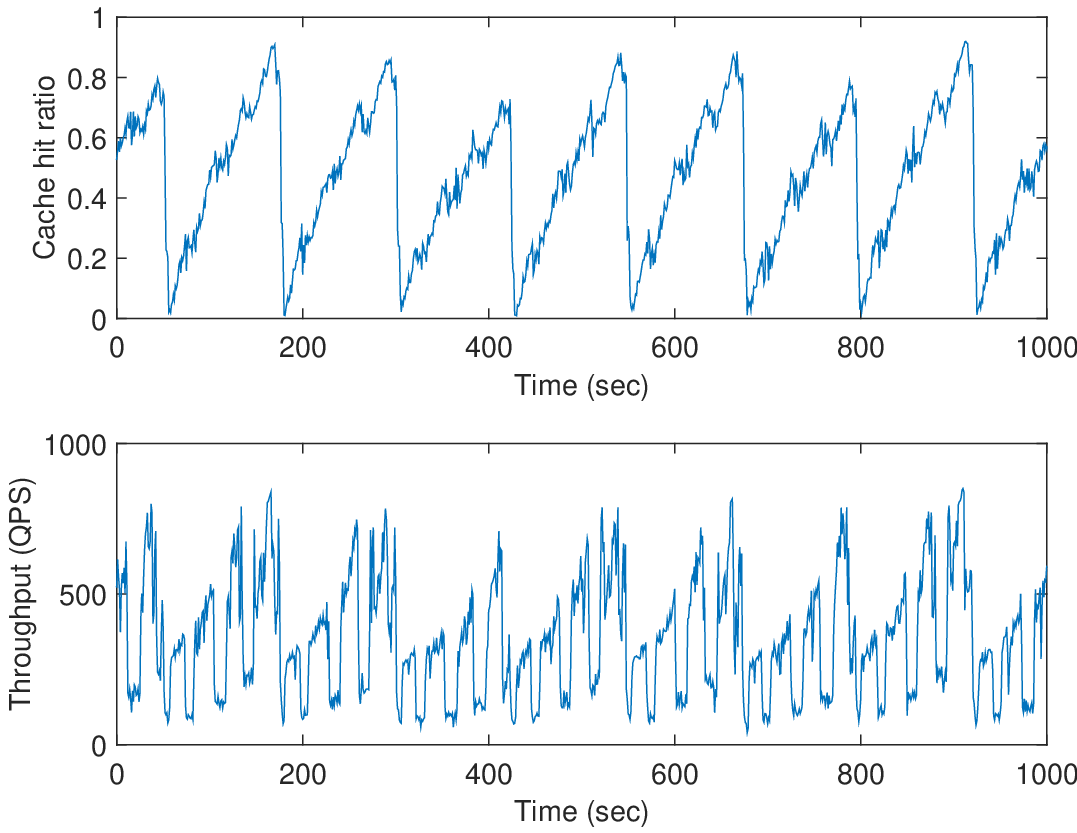}\label{fig:demo-rw}}
}
\vspace{-3mm}
\caption{The failure of caching in LSM-tree}
\label{fig:demo}
\vspace{-5mm}
\end{figure}

Figure \ref{fig:demo} shows the cache hit ratio and random read throughput
for two workloads running on a standard LSM-tree 
(see Section \ref{sec:exp} for the detailed experimental setup in our work). 
With a read-only workload, no compaction is conducted.  As shown in Figure \ref{fig:demo-readonly}, 
the cache hit ratio and read throughput are quite stable during the experiment, 
about 96\% and 3,400 QPS, respectively. 
However, with a 1,000 QPS write throughput, the compaction on the disk becomes intensive. 
Periodically, data in the cache are invalidated and then reloaded from disk due to cold misses. 
As shown in Figure \ref{fig:demo-rw}, 
the cache hit ratio fluctuates between 1\% and 92\%, with an average hit ratio 47\%.  
Accordingly, the read throughput fluctuates between 50 and 850, 
about 10\% of the throughput for the read-only workload. 
In summary, with the frequent compactions caused by intensive writes, 
the cache hit ratio is reduced half and the performance is reduced 90\%. 

\subsection{Existing solutions}




One possible solution is to use a key-value cache instead of a block cache to avoid cache invalidation.  However, LSM-tree relies on the block cache, either managed by OS buffer cache or by LSM-tree itself, to load data from disk. Maintaining another key-value cache would cause more memory consumption and has other non-trivial effort, as shown in \cite{ZWYGLX15}.  
Without addressing the data referencing issues of block cache, 
existing systems either cache more data than the working set in the main memory, 
or replace HDD storage with solid state drives (SSDs), 
as a workaround solution to serve 
random accesses to LSM-tree. 

A typical approach is to hold newly compacted data in memory for a certain time. 
In \cite{AK15}, the authors propose a pragmatic approach for HBase, which uses dedicated servers
to conduct compactions and keeps the compacted data in the main memory of these servers, 
working as a remote cache to serve requests of data nodes. 
However, the frequently accessed data may only account for part of these compacted data. 
Without addressing the structural issue of LSM-tree, this approach may work well 
in a distributed computing system with enough idle resources such as large clusters.

Another approach is to replace HDD storage with SSDs to serve random reads on LSM-tree,
such as RocksDB and others \cite{rocksdb,WSJOLZC2014}. 
However, this approach does not address the compaction-induced cache invalidation issue either. 
Thus, the benefit of DRAM cache, which is several hundreds times faster than SSD, cannot be well utilized. 
Meanwhile, LSM-tree workloads are write-intensive, and the compaction
can further amplify the total write volume by tens to hundreds times. 
These excessive writes not only interfere with random reads, 
causing significant performance degradation \cite{CLZ11};
but would also cause a high number of program/erase cycles 
on SSDs that have a limited endurance. 
Furthermore, the high volume sequential writes and 
the same amount of sequential reads generated by compactions
account for a majority of I/O operations of LSM-tree, 
which are less cost-effective on SSDs than on HDDs. 
Thus, a full SSD solution can be expensive and inefficient 
though it does improve the performance.

\subsection{Our solution and contributions}

In this paper, we propose \emph{dLSM-tree} (dLSM in short), 
in order to enable high speed caching for write intensive data. 
The basic idea of dLSM is to place an on-disk \emph{compaction buffer} between the cache
and the LSM-tree, working as a cushion to minimize the cache invalidations caused by compactions. 
The compaction buffer maintains a consistent view of the frequently compacted data in the underlying LSM-tree, 
but is updated at a much lower rate than the compaction rate. 
As a result, data in the compaction buffer are almost stationary and can be kept in the cache effectively. 


In dLSM, the compaction buffer only contains frequently compacted, 
upper-level data of the underlying LSM-tree, 
typically less than 10\% of the total data size. 
By reusing the previously compacted data, the compaction buffer can be built with
very low overhead. 
Furthermore, the ordered structures in the previously compacted data are kept for efficient key lookup. 
Our construction algorithm ensures that for each level, 
data in the compaction buffer are consistent with those in the underlying
LSM-tree at any given time. 
With less than 1\% additional memory usage, our algorithm can achieve the same lookup 
performance as that of a standard LSM-tree. 


dLSM directs key-value queries to the compaction buffer for frequently compacted data
and underlying LSM-tree for infrequently compacted data, both are resident on the disk. 
Thus, frequently requested objects can be effectively kept 
in a cache of the corresponding disk data for fast access, 
increasing query performance cost-effectively. 
Range queries are directed to the underlying LSM-tree on the disk, 
since the compacted data are efficient to scan in the key order. 
Furthermore, with a high hit ratio in the cache, 
the performance churns caused by the interference between random disk accesses
and sequential writes of compactions are also significantly reduced. 
In short, using a small size of disk space as compaction buffer, 
dLSM achieves a high and stable performance for random reads by serving hot data with effective caching, while
retaining all merits of HDD-based LSM-tree. 

We have implemented a prototype of dLSM-tree based on LevelDB \cite{LevelDB}, 
a widely used LSM-tree library developed by Google, 
and conducted extensive experiments with Yahoo! Cloud Service Benchmark (YCSB). 
The evaluation results show that with a standard DRAM-based cache and HDD storage, 
our dLSM implementation can achieve 
5--8x performance improvement over LSM.  



The roadmap of this paper is as follows.
Section \ref{sec:bg} summarizes the background knowledge of LSM-tree. 
Section \ref{sec:motivation} presents the motivation of our dLSM-tree design.
Section \ref{sec:2phase-compact}, \ref{sec:lookup}, and \ref{sec:discuss} 
present the techniques 
to build dLSM and query dLSM, and the service scope of dLSM. 
We present 
the performance evaluation of dLSM in Section 
\ref{sec:exp}.
Finally, we overview the related work in Section \ref{sec:related} 
and conclude this paper in Section \ref{conclusion}.

%% file: background.tex
\section{Background}
\label{sec:bg}



In an LSM-tree, data are stored on disk in multiple levels of increasing sizes,
each of which is organized as a sorted structure 
for the purpose of efficient lookups. 
Newly arrived data are buffered in memory first and then merged into  
the first level on the disk. When the first level is full, its data will be gradually merged
to the second level, and so on. 
The entire process is a sequence of merges (also known as \emph{compactions} in the literature)
level by level, which are conducted progressively in parallel, called \emph{rolling merges}. 
All levels are sorted separately, thus the key spaces of different levels can overlap.
During a rolling merge, only sequential I/O operations are involved, and the data in each
level are kept sorted all the time.
In this way, data are written to disk in a log fashion, and continuously merged
to keep the sorted structure, which leads to the name \emph{Log-Structured Merge-tree}.


\begin{figure}[t]
\centering
\includegraphics[width=0.5\linewidth]{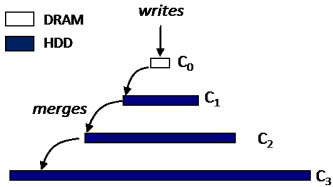}
\vspace{-3mm}
\caption{LSM-tree}
\label{fig:lsm-tree}
\vspace{-5mm}
\end{figure}

Figure~\ref{fig:lsm-tree} shows the structure of an LSM-tree with three on-disk levels. 
Following the notations in~\cite{OCGO96},
the write buffer in main memory is called $C_0$ (level 0),
the first level on the disk is called $C_1$ (level 1), 
the second level is called $C_2$ (level 2), 
and so on.
Denote the number of on-disk levels of an LSM-tree as $k$, and 
denote the sizes of components $C_0, C_1, ..., C_k$ as $S_0,  S_1,  ...,  S_k$, correspondingly. 
We call such an LSM-tree as a $k$-level LSM-tree, or a $k+1$ components LSM-tree as in other literature. 
To minimize the total amount of sequential I/O operations on disk, 
the size ratio $r_i$ between $C_{i}$ and $C_{i-1}$ ($r_i = S_i/S_{i-1}, 1 \leq i \leq k$),  
should be a constant for all levels, denoted as $r$ \cite{OCGO96}. 
We call such an LSM-tree as a \emph{balanced} LSM-tree. 
A small size ratio $r$ corresponds to a high number of levels, $k$. 
Since the key spaces of different levels can overlap, a key lookup may need to access multiple levels.
Considering the high cost of random accesses on HDDs, the size ratio $r$ is usually large 
in order to keep the total number of on-disk levels, $k$, small. 
The size ratio of LevelDB is 10. 

In an LSM-tree, deletions and updates are inserted as special entries into $C_0$, 
and the real deletions or updates
are performed when these entries are merged with their previous versions in a compaction.
Thus, LSM-tree natively supports multi-version concurrency control (MVCC) 
for transactions. 
When a query is initiated, the data blocks to be accessed  
are locked for read-only accesses. 
During the query execution, these blocks remain on disk 
even if they are compacted to new blocks in the next level. 
LSM-tree is also efficient on range queries though often used as a key-value store. 
Since data in each level are sorted, 
a range query just needs to sequentially scan the corresponding key ranges in each level and combines
all results together. 
However, as we will analyze in the following section, 
there is a significant challenge for LSM-tree to be a high performance storage engine of data processing systems. 




%
%

%% file: motivation.tex
\section{Motivation of {dLSM} design}
\label{sec:motivation}

In this section, we will analyze the internals of LSM-tree compactions, 
and show how compactions generate frequent data updates on disk, 
and how these updates interfere with buffer caching, especially for time-sensitive data. 
We then propose the idea of compaction buffer, which works as a cushion between the cache
and the LSM-tree, 
in order to reduce the cache invalidations caused by compactions.  
Finally, we propose the design of dLSM based on this concept.

\subsection{The interference from compactions to buffer caching} 
\label{sec:intf-comp-caching}


\subsubsection{The amplified updates caused by compactions}

Let us consider a balanced LSM-tree with a constant write throughput $w_0$. 
For simplicity, let us assume that each level has the same key space, 
where keys are evenly distributed in the space. 
New data are inserted into $C_0$ at rate $w_0$. 
To keep the size of each level constant, 
data in each level need to be moved out and merged to the next level at the same rate 
as the write throughput, $w_0$. 

\begin{figure}[t]
\centerline{
\includegraphics[width=0.6\linewidth]{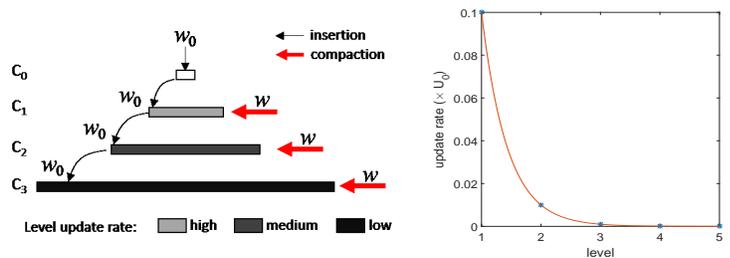}
\hspace{0.015\linewidth}
\includegraphics[height=0.38\linewidth,width=0.44\linewidth]{fig/write-update-rate2.eps}
}
\vspace{-3mm}
\caption{The compaction write rates and level update rates in an LSM-tree}
\label{fig:lsm-write-update}
\vspace{-5mm}
\end{figure}

To keep the LSM-tree balanced, data are moved and merged in a rolling way among multiple levels. 
In each level, only one chunk of data is moved to the next level at a time. 
For each data movement, a compaction is conducted to keep data in the target level sorted. 
A compaction from level $i-1$ to level $i$ ($1 \leq i \leq k$) works as follows. 
First, 
a chunk of data to be moved out from level $i-1$ are read in key order sequentially
from disk into memory. 
Since the size of level $i$ is $r$ times that of level $i-1$, 
for the same key range, data in level $i$ would be $r$ times as data in level $i-1$. 
Thus, $r$ chunks of data in level $i$ are sequentially read from disk into memory too. 
Then the $1+r$ chunks of data from these two levels
are merged in the memory and written to level $i$ on the disk. 

For a $k$-level LSM-tree with write throughput $w_0$, there are $k$ such compactions running concurrently. 
In each level, data are inserted at rate $w_0$, but
the actual write rate on the disk is $w = (1+r) w_0$ due to the compactions, 
as shown on the left side of Figure \ref{fig:lsm-write-update}. 
Thus, the total disk write rate is $k w = k (1+r) w_0$.
This means when a chunk of data is dumped from the write buffer to disk,  
$k(1+r)$ data chunks on the disk will be updated accordingly. 
As a result, the corresponding data kept in the cache, if any, 
have to be invalidated, causing cold misses 
and performance churns as we have presented in Section \ref{sec:intro-problem}.

\subsubsection{Properties of data updates in LSM-tree}

In a $k$-level LSM-tree, although the disk write rates of all levels are the same, $w$,
the size of level $i$ $(1 \leq i \leq k)$ is $r$ times that of its upper level, level $i-1$. 
When level $i$ is completely updated, level $i-1$ has been updated $r$ times. 
Define the \emph{update rate} of level $i$ $(1 \leq i \leq k)$ 
as $U_i = \frac{w}{S_i}$, i.e., the fraction of data updated in this level per unit time. 
For simplicity, denote $U_0 = \frac{w}{S_0}$. We have
\begin{equation}
U_i = \frac{w}{S_i} = \frac{w}{S_0 r^i} = \frac{U_0}{r^i}. 
\label{eq:U}
\end{equation}

Thus, in an LSM-tree, the update rate of a level decreases exponentially from top to bottom,
as shown on the right plot of Figure \ref{fig:lsm-write-update}. 
However, upper-level data are also more recent than lower-level data, 
since with the passage of time, old data
will be compacted to lower levels until the last level in an LSM-tree. 
This means in an LSM-tree, new data are more frequently updated. 

%


In many data systems, such as social media systems like Twitter, 
recently inserted data are much more frequently requested than old ones since
the popularity of objects drops with time quickly \cite{KLPM10}. 
In an LSM-tree, these time-sensitive data are resident in upper levels
and frequently compacted with new data. 
Thus, even though with a high locality of references, the accesses to hot data in these workloads
cannot benefit from the usage of read cache, due to 
the interference from compactions to caching and the structure of LSM-tree.

Although upper-level data are more frequently updated due to compactions, 
they only account for a small fraction of data in the system. 
In a $k$-level LSM-tree, the fraction of data in top $k-1$ levels is
\begin{equation}
\frac{\sum_{i=1}^{k-1}S_i}{\sum_{j=1}^{k}S_j} = \frac{r^{k-1}-1}{r^{k}-1} < \frac{1}{r}. 
\end{equation}

For a typical size ratio used in industry systems, $r=10$, 
top $k-1$ levels data only account for less than 10\% data.

\subsection{dLSM design: Reducing cache invalidations with a compaction buffer}
\label{sec:full-mirror}


\begin{figure}[t]
\centering
\includegraphics[width=0.5\textwidth]{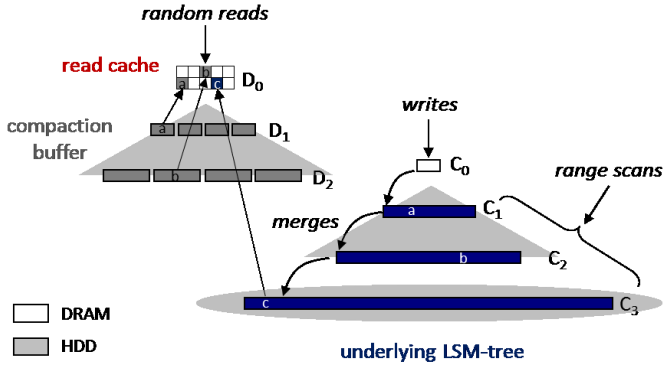}
\vspace{-3mm}
\caption{dLSM-tree architecture}
\label{fig:dLSM}
\vspace{-5mm}
\end{figure}

The interference from compactions to caching in LSM-tree 
is caused by the re-addressing of cached objects in the underlying storage. 
Our analysis has shown that the compactions in top levels of LSM-tree play a major role
for this interference. This motivates us to control and manage the dynamics
of the compactions in top levels of an LSM-tree to achieve effective caching. 
The basic idea is 
to place a \emph{compaction buffer} between the read cache and the LSM-tree, 
working as a cushion to minimize the cache invalidations 
caused by compactions. 
The compaction buffer maintains a separate copy of upper-level data on the disk, 
which is organized in a different way but is consistent with that in the underlying LSM-tree. 
Data in the compaction buffer are updated in a much lower rate than the compaction rate. 
Thus, caching data in the compaction buffer is equivalent to caching the original LSM-tree data, 
but the frequent cache invalidations are avoided and a high hit ratio can be achieved. 
Since the compaction buffer is on disk, 
we call the entire data structure, including read cache, compaction buffer, and the underlying LSM-tree,
\emph{dLSM-tree}, or dLSM in short, as illustrated in Figure \ref{fig:dLSM}.


Figure \ref{fig:dLSM} shows the structure of a three-level dLSM-tree. 
The lower right of the figure is the underlying LSM-tree, which consists of four components
$C_0$, $C_1$, $C_2$, and $C_3$. 
Among them, $C_0$ is the \emph{write buffer} in DRAM to batch writes, 
and $C_1$, $C_2$, and $C_3$ are on-disk levels. 
The upper left is the \emph{read cache}, $D_0$, which is used to 
cache frequently requested data. 
In the middle of the figure, between the read cache and the underlying LSM-tree is
the \emph{compaction buffer}, $D$, which is on disk. 
The compaction buffer has two levels, $D_1$ and $D_2$, 
corresponding to $C_1$ and $C_2$ in the underlying LSM-tree. 
Data in $D_1$ and $D_2$ are consistent with data in $C_1$ and $C_2$, 
but are updated in a low rate to enable effective caching.

In dLSM, random read requests are dispatched to the compaction buffer and the last level of the LSM-tree. 
As shown in Figure \ref{fig:dLSM}, random reads to objects $a$, $b$, and $c$ 
are served from the cache. Among them, $a$ and $b$ are loaded from $D_1$ and $D_2$ of the compaction buffer
instead of $C_1$ and $C_2$ of the underlying LSM-tree, which are frequently updated. 
Object $c$ is loaded from $C_3$ of the LSM-tree directly, since $C_3$ is updated infrequently. 
Range queries are dispatched to the underlying LSM-tree for efficient scan.

%% file: compaction.tex
\section{Building {dLSM}-tree}

\label{sec:2phase-compact}

The compaction buffer needs to be built with low overhead, and 
to keep its key-value data easy to lookup and consistent with those in the underlying LSM-tree. 
We propose the \emph{two-phase compaction} algorithm, which restructures the layout of standard LSM-tree
to prepare data for the compaction buffer, and the \emph{multi-segment cascading} algorithm, 
which builds the compaction buffer with these prepared data and new data dumped from memory. 
The disk space used by compaction buffer is limited and the I/O cost for compaction buffer construction is very low, while data in the compaction buffer are kept ordered for fast lookup. 
Next, we will present the detailed techniques to build dLSM-tree.

\subsection{Two-phase compaction}

\label{sec:2pc}


\begin{figure}[t]
\centerline{
\includegraphics[width=0.99\linewidth]{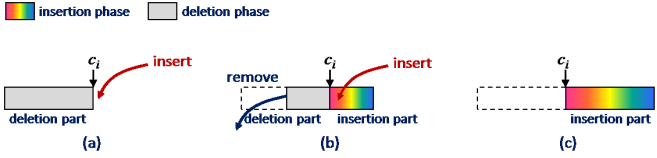}
}
\vspace{-3mm}
\caption{Two-phase compaction of dLSM. (a), (b), and (c) are the beginning, the middle, and the end of a compaction round.}
\label{fig:dlsm-2pc}
\vspace{-2mm}
\end{figure}


\begin{figure}[t]
\includegraphics[width=1\linewidth]{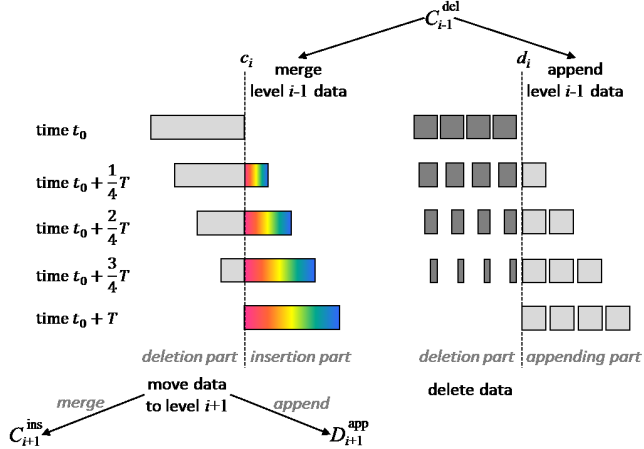}
\vspace{-3mm}
\caption{Data movement procedure of a full compaction round (left: LSM data, right: compaction buffer data)}
\label{fig2pcmsctogether}
\vspace{-5mm}
\end{figure}






In a standard LSM-tree, data in each level are fully sorted. 
As shown on the left side of Figure \ref{fig:lsm-write-update}, 
there are two merges running on level $i$ ($1 \leq i < k$) concurrently: 
a merge from level $i-1$ to level $i$, which inserts data into level $i$; 
and a merge from level $i$ to level $i+1$, which deletes data from level $i$. 
Both insertions and deletions are conducted in key order over the entire level.

In dLSM, the insertions and deletions are conducted in two different parts of the same level. 
As shown in Figure \ref{fig:dlsm-2pc}, 
level $i$ ($1 \leq i < k$) is separated into two parts by the marker $c_i$: a \emph{insertion part}, $C_i^{ins}$, 
where data coming from its upper level are merged into the existing data; and 
a \emph{deletion part}, $C_i^{del}$, 
where data are moved out in key order and merged to the next level. 
Data in both parts are sorted by key order. 
Both the size of $C_i^{del}$ and the size of $C_i^{ins}$ in a level change over time,
however, the sum of the sizes of $C_i^{del}$ and $C_i^{ins}$ is a constant,
corresponding to the level size in a standard LSM-tree. 

Assume initially all data in the level are fully sorted, i.e.,
the deletion part $C_i^{del}$ is full and the insertion part $C_i^{ins}$ is empty,
as shown in Figure \ref{fig:dlsm-2pc}(a). 
New data are inserted into the insertion part by compactions, 
thus the size of the insertion part increases. Meanwhile, 
the size of deletion part decreases because data in this part are moved out 
and merged to the next level (Figure \ref{fig:dlsm-2pc}(b) ). 
Finally, the deletion part becomes empty and
the insertion part becomes full (Figure \ref{fig:dlsm-2pc}(c) ). 
Then the insertion part becomes a deletion part and a new empty insertion part is created, 
forming a loop from Figure \ref{fig:dlsm-2pc}(c) back to Figure \ref{fig:dlsm-2pc}(a).
We call such a procedure a \emph{compaction round} or \emph{merge round}. 

During the period of a compaction round, 
data in a level experience two phases. 
First, when a tuple is put into the insertion part, it enters the \emph{insertion phase}:
in this phase, the location of this tuple on the disk keeps changing due to compactions.
Then after the insertion part becomes a deletion part, it enters the \emph{deletion phase}, 
during which the location of this tuple on the disk keeps unchanged until it is removed from this level. 
In contrast, in standard LSM, the location of a tuple keeps changing before it is merged to the next level. 
Thus, we call this algorithm \emph{two-phase compaction} (2PC). 

Same as in a standard LSM-tree, data in the insertion part are frequently updated and the cached blocks would be invalidated accordingly. 
However, data in the deletion part are just removed sequentially, 
thus the remaining data can still be effectively cached to serve random read requests. 
When the insertion part is full and becomes the deletion part, the entire level is completely sorted. 
dLSM keeps these compacted data as a snapshot of the level on disk (as shown in the dashed line box in Figure \ref{fig:dlsm-2pc}) 
and moves its data to the compaction buffer when they are removed from the underlying LSM-tree. 
In this way, building the compaction buffer makes the least effort. 
We will present the detailed operations in the next subsection. 

\subsection{Building the compaction buffer}

\label{sec:ds}

Figure \ref{fig2pcmsctogether} 
shows the underlying LSM-tree and the compaction buffer during a compaction round side by side. 
The left side of the figure is the underlying LSM-tree, 
and the right side of the figure is the compaction buffer. 
As shown on the right of Figure \ref{fig2pcmsctogether}, data in the compaction buffer are separated into two parts
by the marker $d_i$. The left of the maker is the \emph{deletion part}, $D_i^{del}$, 
corresponding to the deletion part in the underlying LSM-tree; 
the right of the marker is the \emph{appending part}, $D_i^{app}$, 
corresponding to the insertion part in the underlying LSM-tree. 
In each level of the compaction buffer, 
the changes of data are driven by the compaction process of the same level in the underlying LSM-tree, 
thus the two levels have the same running cycle. 

The entire procedure of data movement for a full compaction round is as follows. 
Denote the period of the compaction round of level $i$ as $T$. 
At the beginning of a compaction round, the deletion part $D_i^{del}$ is full 
and the appending part $D_i^{app}$ is empty (Figure \ref{fig2pcmsctogether}, $t = t_0$). 
During the compaction round, data from the deletion part of the upper level, $C_{i-1}^{del}$, 
which are compacted to the insertion part $C_i^{ins}$ in the underlying LSM-tree, 
are moved to the end of appending part $D_i^{app}$ in the compaction buffer (Figure \ref{fig2pcmsctogether}, $t_0 < t < t_0+T$). 
Thus, data in the compaction buffer are always consistent with data in the underlying LSM-tree, except they are uncompacted. 
Since the size of level $i$ is $r$ times that of level $i-1$, there can be up to $r$ sorted tables
in the appending part $D_i^{app}$ of the compaction buffer. 
Meanwhile, in the underlying LSM-tree, data in the deletion part $C_i^{del}$ are moved out in key order and compacted to the next level. 
These data correspond to data in the same key range of the $r$ separately sorted $D_i^{del}$ tables
in the compaction buffer, 
which will be deleted from the disk to keep the level size constant. 
Finally, at the end of the compaction round, the deletion part is empty and the appending part is full 
(Figure \ref{fig2pcmsctogether}, $t = t_0+T$). 
Then the appending part becomes a deletion part and a new appending part is created for the next round of compaction. 
Since data in the level $i$ of the compaction buffer
are constructed by sequentially concatenating $r$ snapshots of 
level $i-1$ in the underlying LSM-tree, 
we call this algorithm \emph{multi-segment cascading} (MSC).

\subsection{Disk I/O and storage cost}
\label{sec:dlsm-io-cost}

With standard LSM compaction, 
a data block from level $i-1$ would be merged with $r$ blocks in level $i$ 
since the size of level $i$ is $r$ times that of level $i-1$. 
With two-phase compaction, the size of $C_i^{ins}$ increases from zero to the full size of level $i$ 
during a merge round, 
so on average a block from $C_{i-1}^{del}$ would be merged with $r/2$ blocks in $C_{i}^{ins}$ during this round. 
Thus, the average amount of I/O operations in two-phase compaction is only about half
of that in standard LSM compaction.
The sequential reads and sequential writes in two-phase compaction are still half-to-half. 

However, the lookup cost in a level with two-phase compaction is greater than that with standard LSM compaction. 
During a two-phase compaction round, the key range of $C_i^{ins}$ part is always the full key space, 
and the key range of $C_i^{del}$ part decreases from the full key space to zero with the removal of its data. 
Thus, for a key-value lookup or range scan, the average number of overlapped ranges is 1.5 
for two-phase compaction, while it is one for standard LSM compaction. 
Nevertheless, 
in both LSM and dLSM, the tradeoff between lookup cost and compaction I/O can be tuned by the number of levels. 
From this perspective, two-phase compaction provides a flexibility for this tuning. 
The last level of dLSM-tree is compacted into a fully sorted table instead of using two-phase compaction,
in the same way as that in standard LSM-tree. 

The I/O cost of multi-segment cascading is very low. 
Level 1 data are built from level 0, which are resident in the memory, 
thus 
need to be written to the compaction buffer on disk. 
Except level 1, dLSM does not conduct any write operation in the compaction buffer. 
Other levels data are constructed from the snapshots of levels in the underlying LSM-tree, 
without any I/O operation.  
Thus, the I/O operations to build the compaction buffer 
is just the sequential writes to dump new data from memory, 
a constant independent of the size of the compaction buffer. 
The storage cost of compaction buffer is also low, since HDD storage is inexpensive
and upper-level data only account for less than 10\% data in the system,
as we have analyzed in Section \ref{sec:motivation}.


Data update rates in the compaction buffer of dLSM are much smaller than
those in LSM. Please refer to Appendix \ref{sec:data-update-rate} for a detailed analysis.

%% file: queryprocess.tex
\section{Query processing}
\label{sec:lookup}

\begin{figure}[t]
\centering
\includegraphics[width=0.49\textwidth]{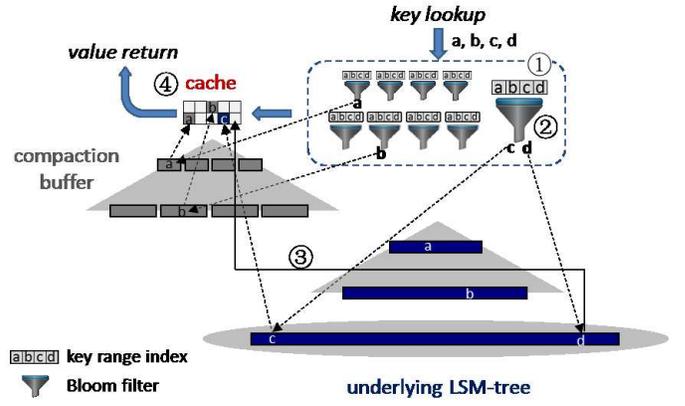}
\vspace{-5mm}
\caption{Work flow of key-value search in dLSM. 
\protect\circled{1} index lookup, 
\protect\circled{2} Bloom filter test,
\protect\circled{3} load data to cache,
\protect\circled{4} serve data from cache. }
\label{fig:dLSM-lookup}
\vspace{-3mm}
\end{figure}

\subsection{Key-value queries}
\label{sec:kv-search}

In dLSM, key-value queries are conducted against 
the compaction buffer and the last level of the underlying LSM-tree, 
which consist of multiple sorted tables with overlapped key ranges. 
Figure \ref{fig:dLSM-lookup} shows the work flow of key-value search in dLSM. 
As shown in the dash line box of Figure \ref{fig:dLSM-lookup}, 
for each sorted table, 
a tree index is built to locate the disk address of the data block whose key range contains the target key. 
In addition, a Bloom filter is built to check whether a key is contained in the table.
A key-value search has four steps. 
The first step is to get the disk addresses of all data blocks 
whose key ranges cover the target key by looking up the index of each table (step \circled{1}). 
Then in the second step, data blocks that do not contain the target key are excluded by Bloom filter tests. 
If there are multiple blocks containing the target key,  
the most recently updated block is selected since it contains the newest version (step \circled{2}). 
In the third step, the data block containing the target key-value object is 
read into the cache from disk, if it is not there yet (step \circled{3}). 
Finally, the target object is served from the cache (step \circled{4}). 

The first two steps of a key-value search are to locate the disk address of the target object, 
for which Bloom filters play an important role. 
Bloom filter is a space-efficient data structure for membership test. 
A negative test means the element is not present in the set. However,
a positive test only means a high probability that the element is present in the set,
with a false positive rate associated. 
Bloom filter has been used in many LSM-tree implementations including LevelDB. 
However, it has been reported that the benefit of Bloom filters becomes diminishing
when the number of tables to be checked is greater than 40 \cite{SSZDA11}. 
As shown in Figure \ref{fig2pcmsctogether}, 
a level in the compaction buffer has up to $r$ sorted tables
in the deletion part and up to $r$ sorted tables in the appending part, with overlapped key ranges. 
In the extreme case, a key can be covered by the key ranges of all these $2r$ tables. 
However, during a compaction round,
the number of tables in the deletion part decreases from $r$ to zero and 
the number of tables in the appending part increases from zero to $r$.
Thus, the average number of overlapped ranges for key lookups 
in a compaction buffer level is $r$, reducing the effectiveness of Bloom filter.

dLSM addresses this issue by allocating a higher number of bits for each element in the filter than LSM. 
To achieve a false positive rate no more than $f$, 
the number of bits per element in the Bloom filter should be at least $\log_2 (\frac{1}{f}) / \ln 2$,
with $\log_2 (\frac{1}{f})$ hash functions in the filter \cite{BM2004}. 
Since a key lookup in the compaction buffer 
needs $r$ filter tests per level, 
the total number of false positives for a level test would be $r$ times that in a standard LSM-tree. 
To achieve the same false positive rate for level membership tests, $f$, 
the false positive rate of each Bloom filter in the compaction buffer should be $f/r$. 
Thus, for Bloom filters in the compaction buffer,
the number of bits per element should be increased to $\log_2 (\frac{r}{f}) / \ln 2$. 

In general, a 1\% false positive rate, 
for which the number of bits per element is 10, is good enough to most applications. 
With the default configuration of LevelDB ($r = 10$), 
the number of bits per element is about 14.4. 
However, since compaction buffer only contains upper-level data in the system, 
the last level of the dLSM-tree, which account for more than 90\% data in the system, 
does not need extra bits for Bloom filters. 
For a three-level LSM-tree, with YCSB benchmark (1 KB key-value size), 
the additional memory consumption for the extra bits of Bloom filters
is less than 1\% the memory usage of a standard LSM-tree, without considering the read cache size.


\subsection{Range queries}

In dLSM, range queries are conducted against the underlying LSM-tree 
in the same way as those in a standard LSM-tree. 
However, data accessed by range queries can also be cached in memory. 
Thus, for key-value queries, an object covered by the compaction buffer may be cached in both
the corresponding read cache and the cache for range queries, 
since it is contained in two different blocks on the disk. 
As a system framework, dLSM does not define the detailed 
cache management strategies, such as cache replacement
and de-duplication policies. 
A simple approach is to check the cache for range queries first. 
If the object is found, 
the block in the compaction buffer containing it will not be loaded into the read cache, $D_0$. 
If the block is already there, it will not be accessed and thus be replaced when $D_0$ is full.

%% file: discuss.tex
\section{The service scope of dLSM}

\label{sec:discuss}

LSM-tree is designed for write-intensive workloads without considering buffer cache in its design space. 
It works well for workloads of write-only and writes with occasional reads or low read throughput. 
dLSM can achieve a comparable performance to LSM-tree on such workloads, 
since the I/O cost of compaction buffer is very low. 
When reads and writes are both intensive, 
dLSM outperforms LSM dramatically, because the compaction buffer 
improves the caching performance significantly. 



For workloads with dynamic read and write throughput, 
dLSM can also boost the overall performance 
by adaptively tuning the speed of compactions and the size of compaction buffer. 
In a live system, 
when the write throughput becomes low, the compaction can still be conducted at a high speed, 
in order to merge the upper-level data to the lowest level quickly. 
We call this \emph{aggressive compaction}. 
When all data are compacted into a single table on the disk, 
both the key-value and range queries would become faster. 
Meanwhile, 
since the compaction buffer becomes empty, the associated resources can be freed. 
However, during the course of aggressive compaction, the read performance of LSM-tree 
would be significantly degraded due to cache invalidations. 
In contrast, with dLSM, a high read performance can be achieved even with high speed compaction. 
In summary, dLSM can serve a general write-intensive workload 
with a high read and write performance at a low cost.

%% file: experiments.tex
\section{Experiments}

\label{sec:exp}

\subsection{Experimental setup}

\label{sec:exp-setup}

Our dLSM-tree implementation is built with LevelDB 1.15 \cite{LevelDB}. 
It implements the two-phase compaction and multi-segment cascading algorithms 
in the LevelDB code framework. 
The implementation of dLSM read cache is based on the LRU block cache of LevelDB. 
We have also implemented the stepped-merge algorithm (SM) \cite{JNSSK97},
which compacts data lazily without keeping the underlying sorted structure of LSM-tree, as a comparison. 
A brief introduction of SM algorithm can be found in Section \ref{sec:related}. 

The hardware system is a storage server running Linux kernel 3.2.0,
which has two quad-core Intel E5354 processors, 8 GB main memory, and
two Seagate hard disk drives (Seagate Cheetah 15K.7, 450GB) configured as RAID0
as the storage for LSM-tree. 
The HDD RAID uses the ext4 file system,
and the I/O queue scheduler is set to \texttt{cfq}.




In our experiments, the size of SSTable file, the basic unit of compaction in LevelDB, is set to 8 MB, 
and the key-value pair size is set to 1 KB. 
The Bloom filter is set to 15-bit per element, 
for data in both compaction buffer and the underlying LSM-tree. 
The write buffer size is set to 100 MB and the total size of data is 100 GB. 
Since the size ratio $r$ is 10, we have three levels on disk. 
The compaction buffer contains data of the first two on-disk levels. 
When the compaction buffer is disabled, the dLSM-tree can work as an LSM-tree with two-phase compaction. 
As pointed out in Section \ref{sec:dlsm-io-cost}, the two-phase compaction
just provides a higher flexibility than the standard LSM compaction for performance tuning. 
Thus, we apply the two-phase compaction in LSM-tree for a fair comparison. 


The workloads in our experiments are based on 
\emph{Yahoo! Cloud Serving Benchmark} (YCSB) \cite{YCSB}, 
which provides a number of workload templates abstracted from real-world applications. 
We adopt the \emph{latest} workload generation model, which characterizes 
the popularity decaying effect over time, and build the following workloads: 
\emph{Latest Uniform}, \emph{Latest Zipfian}, and \emph{Latest RangeHot}. 
Latest Uniform workload characterizes random requests, where the requested tuples are evenly
distributed among recently inserted/updated data. 
Latest Zipfian workload characterizes requests with strong temporal locality, 
where the request frequency of tuples follows the Zipf law, 
and the recently inserted/updated tuples are requested at higher frequencies.
Latest RangeHot workload characterizes requests with strong temporal locality on a given key range, where
the requested tuples are evenly distributed among a key range of recent data, 
about 10\% of the entire key space. 
These workloads are generated in run time with the \texttt{db\_bench} utility provided in the LevelDB package. 
The performance of LSM and dLSM on non-time-sensitive workloads are about the same. 
Due to page limits, we do not present these results. 



\subsection{Performance evaluation}
\label{sec:pe}




\subsubsection{Random read performance}

Our first set of experiments is to evaluate the random read performance of dLSM under intensive writes. 
We set the write throughput to 1,000 QPS, all in updates, 
which are sent via a single thread.  The key-value queries are sent with another thread. 
During the experiment, the cache hit ratio, 
read throughput, and read latency are measured online and dumped to a log file on the disk. 
As a baseline for comparison, we also run the Latest Zipfian workload on a read-only LSM-tree, 
without any write and compaction. 
The size of read cache is set to 500 MB. 
Each experiment is conducted 1,500 seconds after 15 minutes warming up. 

\begin{figure}[t]
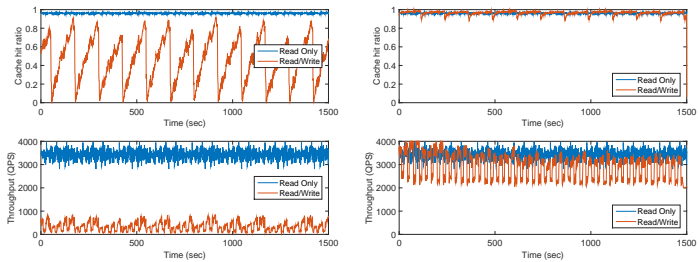

\hspace{-2mm}
\centerline{
\subfigure[LSM]{\includegraphics[width=0.5\linewidth]{fig/demo-rw-lsm_rangehot10_ws1000_2000s-2in1.eps}\label{fig:rangehot-lsm}}
\hspace{0.01\linewidth}
\subfigure[dLSM]{\includegraphics[width=0.5\linewidth]{fig/demo-rw-dlsm_rangehot10_ws1000_2000s-2in1.eps}\label{fig:rangehot-dlsm}}
}
\vspace{-3mm}
\caption{Latest RangeHot workload}
\label{fig:rangehot}
\vspace{-5mm}
\end{figure}

\noindent {\bf Latest RangeHot workload}
Figure \ref{fig:rangehot-lsm} 
and Figure \ref{fig:rangehot-dlsm} 
show the cache hit ratios and read throughputs of the Latest RangeHot workload
running on LSM and dLSM, respectively, with the results of the read-only workload as a comparison. 
As shown in Figure \ref{fig:rangehot-lsm}, 
the cache hit ratio of the read-only workload is very stable, more than 96\% on average,
meaning most frequently accessed data are cached. 
The read throughput is also stable, about 3,400 QPS. 
However, with 1,000 QPS write throughput, 
the cache hit ratio fluctuates periodically between 0 and 90\%. 
This is because when newly arrived data are compacted to the key range of the requested data, 
the corresponding data in the cache will be invalidated immediately, causing cold misses. 
After that, the missed data will be loaded from the disk into the read cache gradually, 
and the cache hit ratio increases accordingly. 
At the same time, the read throughput also fluctuates periodically. 
Overall, the average hit ratio of the LSM-tree with intensive writes 
is only about half of the read-only LSM-tree,
and the read throughput is only about 1/10 of the read-only LSM-tree. 

However, as shown in Figure \ref{fig:rangehot-dlsm}, the fluctuation of cache hit ratio in dLSM
is much smaller.  This is because in the compaction buffer, newly arrived data 
will be written to disk directly without being compacted to the existing data stored on the disk. 
The drops of cache hit ratio are caused by the new data, which are much smaller than
the data involved in a compaction. 
This is why the cache hit ratio in Figure \ref{fig:rangehot-dlsm}
has the same fluctuation period as that in Figure \ref{fig:rangehot-lsm}, but the amplitude is much smaller. 
The overall cache hit ratio is about 96\%, very close to the read-only workload. 
Correspondingly, the read throughput of dLSM is also significantly increased, about 8.5 times
the read throughput of LSM-tree. 
There are still churns in the read throughput, due to the low random I/O rate of HDDs
and the I/O contention from compactions. 
This could be improved by using a larger cache to reduce cache misses
and using another thread to serve other requests from memory 
when the current request is blocked on disk access.

The average latency of random read queries in dLSM is only 1/3 the latency of LSM. 
The random read latency in other experiments are similar. 
Due to page limits, we do not present the related figures.


\begin{figure}[t]
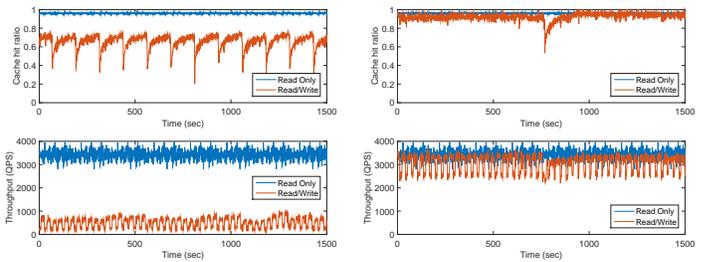

\centerline{
\subfigure[LSM]{\includegraphics[width=0.5\linewidth]{fig/demo-rw-lsm_zipfian0.99_ws1000_2000s.eps}\label{fig:zipf-lsm}}
\hspace{0.01\linewidth}
\subfigure[dLSM]{\includegraphics[width=0.5\linewidth]{fig/demo-rw-dlsm_zipfian0.99_ws1000_2000s-2in1.eps}\label{fig:zipf-dlsm}}
}
\vspace{-3mm}
\caption{Latest Zipfian workload}
\label{fig:zipf}
\vspace{-1mm}
\end{figure}

\begin{figure}[t]
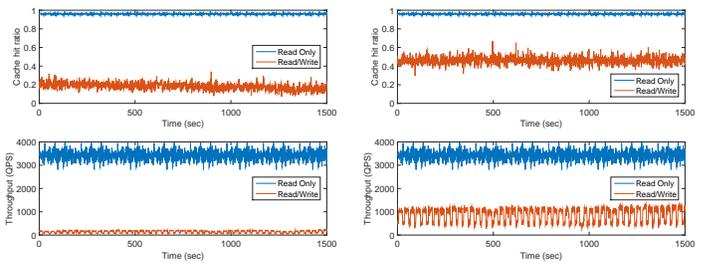

\centerline{
\subfigure[LSM]{\includegraphics[width=0.5\linewidth]{fig/demo-rw-lsm_uniform10hot_ws1000_2000s.eps}\label{fig:uniform-lsm}}
\hspace{0.01\linewidth}
\subfigure[dLSM]{\includegraphics[width=0.5\linewidth]{fig/demo-rw-dlsm_uniformhot10_ws1000_2000s-2in1.eps}\label{fig:uniform-dlsm}}
}
\vspace{-3mm}
\caption{Latest Uniform workload}
\label{fig:uniform}
\vspace{-5mm}
\end{figure}

\noindent {\bf Latest Zipfian workload}
Figure \ref{fig:zipf} shows the cache hit ratios and read throughputs of the Latest Zipfian workload,
running on LSM and dLSM, respectively, 
with corresponding results for the read-only workload as a comparison. 
As shown in Figure \ref{fig:zipf-lsm}, the cache hit ratio also fluctuates periodically, which
is similar to that in the Latest RangeHot workload. 
However, when the cache hit ratio drops to the minimum, 
it will then increase sharply. This is because in the Latest Zipfian workload, a small fraction
of tuples account for the majority of accesses. 
When a frequently requested tuple is reloaded into the cache, the cache hit ratio would
increase significantly. However, the average cache hit ratio is only 65\%, still much lower than 
the read-only workload. 
The read throughput has the similar pattern, with an average value 498 QPS. 

In contrast, as shown in Figure \ref{fig:zipf-dlsm}, the fluctuation period of cache hit ratio in dLSM is
much longer than that in LSM, due to the low update rate of data in the compaction buffer. 
Different from the Latest RangeHot workload, new updates in the compaction buffer
would not affect the cache hit ratio unless they are for frequently requested data. 
However, when these hot data are invalidated due to compaction, the cache hit ratio
will drop sharply. The average cache hit ratio in dLSM is 91\%, and the average throughput is 2900 QPS,
about 5.8 times the throughput of LSM. 

\noindent {\bf Latest Uniform workload}
Figure \ref{fig:uniform} shows the cache hit ratios and read throughputs of the Latest Uniform workload,
running on LSM and dLSM, respectively, 
with corresponding results for read-only workload as a comparison. 
As shown in Figure \ref{fig:uniform-lsm}, the cache hit ratio changes quite smoothly. 
The reason is that in a uniform workload, each tuple has the same probability to be accessed. 
Thus, a compaction can only affect a small fraction of data cached in the read cache. 
However, due to the high frequency of compactions, the cache hit ratio will be 
repressed to a small value, about 0.2 as shown in Figure \ref{fig:uniform-lsm},
and the read throughput is less than 150 QPS. 
In contrast, as shown in Figure \ref{fig:uniform-dlsm}, due to the small data update rate 
in the compaction buffer, the cache hit ratio of dLSM is about 46\%, more than two times that of LSM. 
The average throughput of dLSM is about 866 QPS. 

In summary, our experiments show that dLSM achieves a much higher cache hit ratio (can be two times) and 
a much higher read throughput (can be 5 -- 8 times) than LSM
on write-intensive data of different workloads. 

\subsubsection{Bloom filter and range query performance}

\begin{figure}[t]
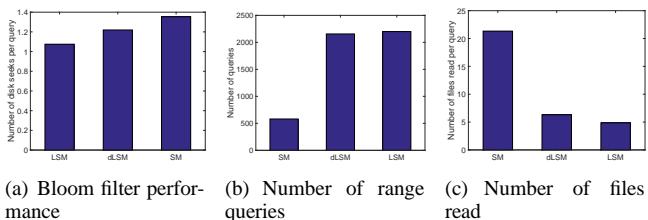

\centerline{
\subfigure[Bloom filter performance]{\includegraphics[width=0.3\linewidth]{fig/number-reads-per-qu.eps}\label{fig:bf}}
\hspace{0.2mm}
\subfigure[Number of range queries]{\includegraphics[width=0.3\linewidth]{fig/sm_lsm_range_qps.eps}\label{fig:range-qps}}
\hspace{0.2mm}
\subfigure[Number of files read]{\includegraphics[width=0.3\linewidth]{fig/sm_lsm_range_files.eps}\label{fig:range-files}}
}
\vspace{-3mm}
\caption{Performance of Bloom filters and range queries}
\label{fig:mb-ssd-read}
\vspace{-5.5mm}
\end{figure}

We have also evaluated the Bloom filter and range query performance of dLSM,
compared with those of LSM and SM. 
Figure \ref{fig:bf} shows the average number of disk accesses to read a tuple directly
from disk, for LSM-, dLSM-, and SM-tree 
with the same size of Bloom filter (15 bits per element). 
SM has the highest number of disk accesses due to the multiple overlapped tables in each level. 
dLSM has a higher number of disk accesses than LSM, but smaller than SM. 
However, as shown in Figures \ref{fig:rangehot}, \ref{fig:zipf}, and \ref{fig:uniform},
the overall random read performance of dLSM is much higher than LSM. 

In order to evaluate the range query performance of dLSM under intensive writes, 
we use the YCSB workload E generator to generate random range queries, 
with 10 MB data for each query, and conduct experiments as follows. 
Same as the first set of experiments, we set the write throughput to 1,000 QPS, 
which is sent via a single thread.  The range queries are sent in another thread. 
The experiments are conducted on LSM-, dLSM-, and SM-trees, each for 500 seconds, respectively.

Figure \ref{fig:range-qps} shows the number of queries executed during these experiments. 
dLSM achieves nearly the same number of queries as LSM. 
In contrast, the query rate in SM is only about 1/4 of that in LSM. 
Figure \ref{fig:range-files} shows the number of SSTable files scanned during a range query, 
where each scan needs a seek. 
The number of files a query needs to scan in dLSM and LSM are about the same, 
since both are based on two-phase compaction. 
SM has the worst range query performance because data in each level are stored 
in multiple separately sorted tables. 
On average, a range query on SM-tree needs to scan 4.4 times files than those 
to be scanned on LSM-tree. 
For efficient range queries, 
the compacted structure of the underlying LSM-tree must be retained.

%
%
%
%

\subsubsection{Performance of dLSM on dynamic workloads}

\begin{figure}[t]
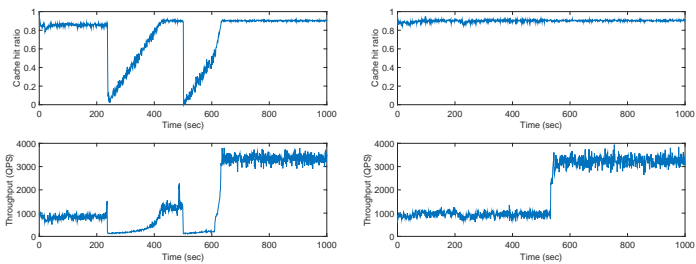

\centerline{
\subfigure[LSM]{\includegraphics[width=0.5\linewidth]{fig/adaptive-lsm.eps}\label{fig:adapt-lsm-read}}
\hspace{0.01\linewidth}
\subfigure[dLSM]{\includegraphics[width=0.5\linewidth]{fig/adaptive-dlsm.eps}\label{fig:adapt-dlsm-read}}
}
\vspace{-3mm}
\caption{Random read performance on LSM and dLSM}
\label{fig:adapt}
\vspace{-1mm}
\end{figure}

\begin{figure}[t]
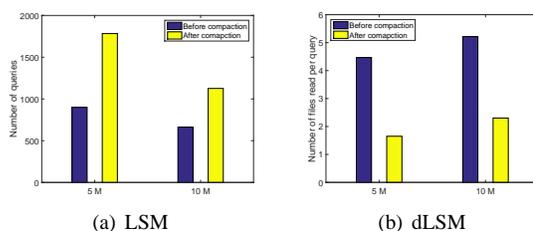

\centerline{
\subfigure[LSM]{\includegraphics[width=0.37\linewidth]{fig/adapt_range_qps.eps}\label{fig:adapt-range-qps}}
\hspace{0.05\linewidth}
\subfigure[dLSM]{\includegraphics[width=0.35\linewidth]{fig/adapt_num_range_files.eps}\label{fig:adapt-num-qu-files}}
}
\vspace{-3mm}
\caption{Range query performance before and after the full compaction}
\label{fig:adapt-range}
\vspace{-5.5mm}
\end{figure}

In order to evaluate the performance of dLSM on serving dynamic workloads, 
first, we set the write rate as zero, but let the compactions continue to run aggressively
until the upper-level data are fully merged to the last level, for both LSM- and dLSM-trees. 
During these compactions, we run the Latest RangeHot workload and measure the random read performance. 

Figure \ref{fig:adapt-lsm-read} and \ref{fig:adapt-dlsm-read} show the cache hit ratio and read throughput 
for LSM- and dLSM-trees, respectively. 
During the first 200 seconds, the performance of LSM and dLSM are both stable,
with very close cache hit ratio and read throughput, because the requested data are not being compacted yet. 
However, when the requested data are being merged to the next level
(about 200 seconds after the measurement)
the cache hit ratio and read throughput of the LSM-tree drop sharply, 
and then gradually increase to its previous peaks in a long time (about 400 seconds). 
In contrast, with the help of the compaction buffer, the cache hit ratio
and read throughput of the dLSM-tree keep unchanged. 
When the first level has been completely compacted to the second level (about 600 seconds after the measurement), 
we paused the compaction temporarily and found 
the performance of the LSM- and dLSM-trees become the same. 
Overall, during the compaction, 
the average cache hit ratio and read throughput of dLSM are 41\% and
91\% higher than those of LSM. 

After the compaction, we run the range query workloads with 5 M and 10 M data in each query, 
on the LSM- and dLSM-tree without other operations, 
each for 100 seconds. 
As shown in Figure \ref{fig:adapt-range-qps}, after a full compaction, the QPS of range queries on 
the LSM/dLSM tree increase 80\% to 120\%. 
However, for LSM-tree, 
this performance gain is at the cost of performance degradation of random reads
during the compaction. 
In contrast, with the compaction buffer, the random read performance of dLSM
is still stable and high during the compaction.
Figure \ref{fig:adapt-num-qu-files} shows the average number of SSTable files scanned in each query.

%% file: related.tex
\section{Other related work}
\label{sec:related}

Stepped-merge (SM) is proposed to balance the tradeoff between compaction I/O and search cost
for data warehouses \cite{JNSSK97}. 
Similar to LSM-tree, SM also organizes data into a multi-level structure of 
exponentially increasing sizes and compacts data with sequential I/Os. 
However, data in a level are only compacted together when they are moved to the next level. 
Thus, the amount of compactions and the corresponding cache invalidations can be reduced significantly. 
However, for a $k$-level SM-tree with size ratio $r$, 
each level would have $r$ sorted tables, whose key ranges can overlap. 
As a result, the range query performance of SM is poor and the cost of key-value lookup is high.
In contrast, dLSM retains the compacted data structure of LSM for efficient range queries
and uses large Bloom filters for high performance key-value search. 


In order to reduce the compaction cost, 
some production systems only compact data partially in run time, 
and run a full compaction during system idle time. 
In HBase, the former is called \emph{minor compaction}, while
the latter is called \emph{major compaction}. 
However, 
disabling major compaction during run time mainly reduces the compaction of old data.
These old data correspond to the last level data in a standard LSM-tree, 
whose update rates are much lower than new data in upper levels, 
and are also less frequently accessed than new data. 
Thus, unlike SM, this approach cannot avoid the interference from compactions to caching. 
In practice, HBase still suffers low read performance during intensive writes \cite{AK15}. 
In contrast, the compaction buffer of dLSM reduces the update rates of new data, thus enabling effective caching. 


bLSM separates the first on-disk level into two parts, $C_1$ and $C_1'$ \cite{blsm2012}. 
The corresponding compaction method is similar to the two-phase compaction in dLSM, 
but is only applied to the first level. The purpose of this separation is 
to synchronize the merge from $C_0$ to $C_1$ and that from $C_1'$ to $C_2$, 
in order to avoid write pause while keeping the level size constant strictly. 
However, this hard constraint of level size may limit the write throughput. 
In LevelDB, the level sizes can be changed in a flexible range to slow down the compactions in order to
avoid write pause, and the structure of LSM-tree will be re-balanced when the write rate is low. 
dLSM adopts the merge scheduling method used in LevelDB. 

A FD-tree is proposed in \cite{LHYLY2010} for data indexing on SSDs without random writes, 
which has a similar idea to LSM-tree 
and is enhanced with fractional cascading technique for a low memory usage. 
However, a lookup needs multiple disk accesses if the object is not found in the first level. 
VT-tree~\cite{SSMASZ13} is an extension of LSM-tree to avoid unnecessary merges for presorted data. 
In \cite{WSJOLZC2014}, the internal parallelism among flash channels is exploited to improve SSD-based LSM-trees. 
\cite{GBHK15} proposes cLSM, an LSM-tree supporting scalable concurrency with multicore processors. 
Different from these studies, our dLSM is a low cost general solution to solve the fundamental problem
of data caching in LSM-tree.

%% file: conclusion.tex
\section{Conclusion}
\label{conclusion}


The original LSM-tree design is optimized for write-intensive workloads, which can cause high cache misses due to compaction-induced cache invalidations. We propose dLSM to address this structural issue, which uses an on-disk compaction buffer to slow down the update rate of frequently compacted data in LSM-tree for effective caching, retaining the underlying structure of LSM-tree for efficient compactions and range queries. 
The service scope of dLSM is applicable to workloads of both intensive reads and intensive writes and dynamic workloads due to its low overhead.  


%% file: appendix-data-update-rate.tex
\section{Data update rate of dLSM}

\label{sec:data-update-rate}


In dLSM-tree, data are appended to the end of each level in the compaction buffer
with multi-segment cascading.  
Different from the LSM compaction, when new data are inserted into the compaction buffer, 
existing data are neither reorganized nor relocated. 
Instead, for level $i$ ($1 \leq i < k$) in the compaction buffer, 
new data are appended to $D_i^{app}$ and old data are removed from $D_i^{del}$, 
both at rate $w_0$, the write throughput of the system. 
Thus, 
the update rate of level $i$ in the compaction buffer, $U^d_i$, is
\begin{equation}
U^d_i = \frac{w_0}{S_i} = \frac{1}{1+r} \frac{w}{S_i} = \frac{U_i}{1+r}, 
\end{equation}
where $U_i$ is the update rate of level $i$ in a standard LSM-tree. 
For typical LSM-tree systems, $r=10$, thus the update rate in the compaction buffer is less than 10\% that in standard LSM-tree.

In a dLSM-tree, the update rate of data in the last level,
$U^d_k$, is the same as that of a standard LSM-tree, since they are compacted in the same way.
We have
\begin{equation}
U^d_{k} = U_{k} = \frac{w}{S_k} = \frac{1+r}{r} \frac{w_0}{S_{k-1}} 
\approx \frac{w_0}{S_{k-1}} = U^d_{k-1},
\end{equation}
where $U^d_{k-1}$ is the update rate of the lowest level in the compaction buffer. 
In general, the update rate of last level data in dLSM 
is already small enough and thus does not need to be kept in the compaction buffer.